\documentclass{llncs} 
\usepackage{amsmath,amssymb}

\newcommand{\remove}[1]{}

\newcommand{\C}{{\mathbb C}}
\newcommand{\F}{{\mathbb F}}
\newcommand{\Z}{{\mathbb Z}}
\newcommand{\ord}{{\mathcal O}}
\newcommand{\e}{{\rm e}}
\newcommand{\eps}{\epsilon}
\newcommand{\poly}{{\rm poly}}
\newcommand{\polylog}{{\rm polylog}}
\newcommand{\re}{{\rm Re}\,}
\newcommand{\U}{\textrm{U}}
\newcommand{\id}{\mathbf{1}}

\newcommand{\ve}{\left( \! \begin{array}{c}}
\newcommand{\ctor}{\end{array} \! \right)}
\newcommand{\mat}{\left( \! \begin{array}{rr}}
\newcommand{\rix}{\end{array} \! \right)}
\newcommand{\norm}[1]{\left\| #1 \right\|}
\newcommand{\abs}[1]{\left| #1 \right|}

\newcommand{\Ind}{\textrm{Ind}}
\newcommand{\rank}{\textbf{rk}\;}

\begin{document}

\title{The Hidden Subgroup Problem in Affine Groups: 
  Basis Selection in Fourier Sampling}


\author{Cristopher Moore\inst{1} \and Daniel Rockmore\inst{2} \and
  Alexander Russell\inst{3} \and Leonard J. Schulman\inst{4}}

\institute{University of New Mexico, \email{moore@cs.unm.edu} \and
  Dartmouth College, \email{rockmore@cs.dartmouth.edu} \and University
  of Connecticut, \email{acr@cse.uconn.edu} \and California Institute
  of Technology, \email{schulman@caltech.edu}}

\maketitle

\begin{abstract}
  Many quantum algorithms, including Shor's celebrated factoring and
  discrete log algorithms, proceed by reduction to a \emph{hidden
    subgroup problem}, in which a subgroup $H$ of a group $G$ must be
  determined from a quantum state $\psi$ uniformly supported on a left
  coset of $H$. These hidden subgroup problems are then solved by
  \emph{Fourier sampling}: the quantum Fourier transform of $\psi$ is
  computed and measured. When the underlying group is non-Abelian, two
  important variants of the Fourier sampling paradigm have been
  identified: the \emph{weak standard method}, where only
  representation \emph{names} are measured, and the \emph{strong
    standard method}, where full measurement occurs. It has remained
  open whether the strong standard method is indeed stronger, that is,
  whether there are hidden subgroups that can be reconstructed via the
  strong method but \emph{not} by the weak, or any other known, method.
  
  In this article, we settle this question in the affirmative. We show
  that hidden subgroups of semidirect products of the form $\Z_q
  \ltimes \Z_p$, where $q \mid (p-1)$ and $q = p/ \polylog(p)$, can be
  efficiently determined by the strong standard method. Furthermore,
  the weak standard method and the ``forgetful'' Abelian method are
  insufficient for these groups.  We extend this to an
  information-theoretic solution for the hidden subgroup problem over
  the groups $\Z_q \ltimes \Z_p$ where $q \mid (p-1)$ and, in
  particular, the Affine groups $A_p$. Finally, we prove a closure
  property for the class of groups over which the hidden subgroup
  problem can be solved efficiently.
  
  \textbf{Submission Track: A}

\end{abstract}

\section{The Hidden Subgroup Problem}

Simon's algorithm for the ``XOR-mask'' oracle problem \cite{Simon97}
and Shor's factoring algorithm \cite{Shor97} determine an unknown
(``hidden'') subgroup $H$ of a given group $G$ in the following way.
\begin{description}
\item[Step 1.] Prepare two registers, the first in a uniform
  superposition over the elements of a group $G$ and the second with
  the value zero, yielding the state $\psi = c_G \cdot \sum_{g \in G} |g\rangle  \otimes
  |0\rangle$, where $c_G = 1/ \sqrt{|G|}$.
\item[Step 2.] Calculate a (classical polynomial-time) function $F$
  defined on $G$ and XOR it with the second register.  This entangles
  the two registers and results in the state $\psi = c_G \cdot \sum_{g \in G} |g\rangle
  \otimes |F(g)\rangle$.
\item[Step 3.] Measure the second register.  This produces a uniform
  superposition over one of $F$'s level sets, i.e., the set of group
  elements $g$ for which $F(g)$ takes a particular value $F_0$.  If
  the level sets of $F$ are the cosets of $H$, this puts the first
  register in a uniform distribution over superpositions on one of
  those cosets, namely $cH$ where $F(c)=F_0$.  Moreover, it
  disentangles the two registers, resulting in the state $\psi = (1/
  \sqrt{|H|}) \; \sum_{h \in H} |cH\rangle \otimes |F_0\rangle$.
\end{description}
Write the amplitudes of the basis states in the first register as the
function
\begin{equation}
  \label{eq:super}
  f(g) = \begin{cases}
    1/ \sqrt{|H|} & \text{if}\;g \in cH,\\
    0 & \text{otherwise.}
  \end{cases}
\end{equation}
The approach taken by Simon and Shor is to perform {\em Fourier
  Sampling} \cite{BernsteinV93}: carry out a quantum Fourier transform
on $f$, and measure the result.
  
In Simon's case, the ``ambient'' group $G$, over which the Fourier
transform is performed, is $\Z_2^n$ and $H$ is a subgroup of index
$2$. In Shor's case (factoring), $G$ is the cyclic group $\Z_n^*$
where $n$ is the number we wish to factor, $F(x) = r^x \bmod n$ for a
random $r < n$, $H$ is the subgroup of $\Z_n^*$ of index order$(r)$,
and the Fourier transform is the familiar Abelian one.  (However since
$|\Z_n^*|$ is unknown, the above algorithm is actually performed over
$\Z_q$ where $q$ is polynomially bounded by $n$; see \cite{Shor97} or
\cite{HalesH99,HalesH00}.)  To solve the elusive \textsc{Graph
  Automorphism} problem, on the other hand, it would be sufficient to
solve the HSP over the permutation group $S_n$; see, e.g.,
Jozsa~\cite{Jozsa00} for a review.  It is partly for this reason that
the non-Abelian HSP has remained such an active area of quantum
algorithms research.

In general, we will say that the HSP for a family of groups has a {\em
  Fourier sampling} algorithm if a procedure similar to that outlined
above works. Specifically, the algorithm prepares a superposition of
the form~\eqref{eq:super}, computes its (quantum) Fourier transform,
and measures the result in a basis of its choice.  After a polynomial
number of such trials, a polynomial amount of classical computation,
and, perhaps, a polynomial number of classical queries to the function
$F$ to confirm the result, the algorithm produces a set of generators
for the subgroup $H$ with high probability.

Since we are typically interested in exponentially large groups, we
will take the size of our input to be $n = \log |G|$.  Thus
``polynomial'' means polylogarithmic in the size of the group.

\paragraph{History and Context.} Though a number of interesting
results have been obtained on the non-Abelian HSP, the groups for
which efficient solutions are known remain woefully few and sporadic.
On the positive side, Roetteler and Beth~\cite{RoettelerB98} give an
algorithm for the wreath product $\Z_2^k\; \wr\; \Z_2$.  Ivanyos,
Magniez, and Santha~\cite{IvanyosMS01} extend this to the more general
case of semidirect products $K \ltimes \Z_2^k$ where $K$ is of
polynomial size, and also give an algorithm for groups whose
commutator subgroup is of polynomial size.  Friedl, Ivanyos, Magniez,
Santha and Sen solve a problem they call Hidden Translation, and thus
generalize this further to what they call ``smoothly solvable''
groups: these are solvable groups whose derived series is of constant
length and whose Abelian factor groups are each the direct product of
an Abelian group of bounded exponent and one of polynomial
size~\cite{FriedlIMSS02}.

In another vein, Ettinger and H{\o}yer~\cite{EttingerH98} show that
the HSP is solvable for the dihedral groups in an
information-theoretic sense; namely, a finite number of quantum
queries to the function oracle gives enough information to reconstruct
the subgroup, but the best known reconstruction algorithm takes
exponential time.  More generally, Ettinger, H{\o}yer and
Knill~\cite{EttingerHK99} show that for {\em arbitrary} groups the
HSP can be solved information-theoretically with a finite number of
quantum queries, but do not give an explicit set of measurements to do
so.

Our current understanding, then, divides groups in three classes
\begin{description}
\label{classification}
\item[I. Fully Reconstructible.] Subgroups of a family of groups
  $\mathbf{G} = \{ G_i \}$ are \emph{fully reconstructible} if the HSP
  can be solved with high probability by a quantum circuit of size
  polynomial in $\log |G_i|$.
\item[II. Measurement Reconstructible.] Subgroups of a family of
  groups $\mathbf{G} = \{ G_i \}$ are \emph{measurement reconstructible}
  if the solution to the HSP for $G_i$ is determined
  information-theoretically by the fully measured result of a quantum
  circuit of size polynomial in $\log |G_i|$.
\item[III. Query Reconstructible.] Subgroups of a family of groups
  $\mathbf{G} = \{ G_i \}$ are \emph{query reconstructible} if the
  solution to the HSP for $G_i$ is determined by the quantum state
  resulting from a quantum circuit of polynomial size in $\log |G_i|$,
  in the sense that there is a POVM that yields the subgroup $H$ with
  constant probability. (Note that there is no guarantee that this
  POVM can be implemented by a small quantum circuit.)
\end{description}
In each case, the quantum circuit has oracle access to a function $f :
G \to S$, for some set $S$, with the property that $f$ is constant on
each left coset of a subgroup $H$, and distinct on distinct cosets.

In this language, then, the result of \cite{EttingerHK99} shows that
subgroups of arbitrary groups are query reconstructible, whereas it is
known that subgroups of Abelian groups are in fact fully
reconstructible.  The other work cited above has labored to place
specific families of (non-Abelian) groups into the more
algorithmically meaningful classes I and II above.

All the above results use Abelian Fourier analysis, even in the cases
in which the groups of interest are non-Abelian; it turns out that
each of these groups are ``close enough'' to Abelian that a
``forgetful'' Abelian Fourier analysis, which treats the groups as
though their multiplication rule was commutative, suffices to detect
subgroups.  Nevertheless, as we shall see, there are situations in
which Abelian Fourier analysis will not suffice and, instead, the full
power of the non-Abelian Fourier analysis associated with the group is
required.

Fourier analysis over a finite Abelian group $A$ proceeds by
expressing a function $f : A \to \C$ as a linear combination of special
functions $\chi: A \to \C$ which are \emph{homomorphisms} of $A$ into $\C$.
If $A = \Z_p$, for example, the homomorphisms from $A$ to $\C$ are
exactly the familiar functions $\chi_t: z \mapsto e^{2\pi i tz/p} \equiv \omega_p^{tz}$ and
any function $f : A \to \C$ can be uniquely expressed as a linear
combination of these $\chi_t$; this change of basis is precisely the
Fourier transform. When $G$ is a non-Abelian group, however, this same
procedure cannot work: in particular, there are not enough
homomorphisms of $G$ into $\C$ to even span the space of all
$\C$-valued functions on $G$. The representation theory of finite
groups constructs the objects which can be used in place of the
$\C$-valued homomorphisms above to develop a satisfactory theory of
Fourier analysis over general groups.
See~\cite{Serre77,FultonH91} for treatments of non-Abelian
Fourier analysis and representation theory.  In this general setting
Fourier transforms are matrix-valued and our Fourier sampling
algorithm might measure not just which representation we are in, but
also the row and column. See Appendix~\ref{appendix:Fourier} for more
discussion.

Along these lines, Hallgren, Russell, and Ta-Shma~\cite{HallgrenRT00}
showed that measuring the names of representations alone --- the
\emph{weak standard method} in the terminology of~\cite{GrigniSVV01}
--- can reconstruct normal subgroups (and thus solve the HSP for
Hamiltonian groups, all of whose subgroups are normal).  More
generally, they show how to reconstruct the {\em normal core} of a
subgroup, i.e.\ the intersection of all its conjugates.  On the other
hand, they show that this is insufficient to solve the Graph
Automorphism problem, since even in an information-theoretic sense
this method cannot distinguish between the trivial subgroup of $S_n$
and most subgroups of order 2.

Grigni, Schulman, Vazirani and Vazirani~\cite{GrigniSVV01} showed that
trivial and non-trivial subgroups are still information-theoretically
indistinguishable, even if we do measure the rows and columns of the
representation, under the assumption that a random basis is used for
each representation. In other words, even the \emph{strong standard
method,} in which rows and columns are measured, cannot solve Graph
Automorphism unless there exist bases for the representations of $S_n$
with very special computational 
properties. (They also point out that since we can
reconstruct normal subgroups, we can also solve the HSP for groups
where the intersection of all normalizers (the Baer norm) has
small index.)

\paragraph{Contributions of this paper.}
An important open question, then, is whether there are cases in which
the \emph{strong standard method} offers any advantage over a simple
Abelian transform or the \emph{weak standard method}.  In this paper,
we settle this question in the affirmative.  Our results deal
primarily with semidirect products of the form $\Z_q \ltimes \Z_p$, the
so-called {\em $q$-hedral} groups, including the {\em affine} group
$A_p \cong \Z_p^* \ltimes \Z_p$.  We show the following:
\begin{theorem}
  \label{thm:hsp}
  Let $p$ and $q$ be prime with $q = (p-1)/\polylog(p)$.  Then
  subgroups of $\Z_q \ltimes \Z_p$ are fully reconstructible.
\end{theorem}
More generally, we define the {\em Hidden Conjugate Problem} as
follows: given a group $G$, a non-normal subgroup $H$, and a function
which is promised to be constant on the cosets of some conjugate
$bHb^{-1}$ of $H$, identify $b$. We adopt the above classification
(fully/ measurement/ query) for this problem in the natural way. Then
we also show that
\begin{theorem} 
  \label{thm:hcp}
  Let $p$ be prime and $q$ a divisor of $p-1$.  Then the hidden
  conjugates of $H$ in $G = \Z_q \ltimes \Z_p$ are fully
  reconstructible if $H$ has index $\polylog(p)$.
\end{theorem}
\begin{sloppypar}
Moreover, our algorithms in Theorems~\ref{thm:hsp} and \ref{thm:hcp}
rely crucially on the high-dimensional representations of $\Z_q
\ltimes \Z_p$, and we show that Abelian methods (in other words,
treating the group as a direct product rather than a semidirect one)
do not suffice.
\end{sloppypar}

We also generalize the results of Ettinger and H{\o}yer on the dihedral
group to the \emph{$q$-hedral} groups:
\begin{theorem}
\label{thm:qhedralhcp}
Let $p$ be prime and $q$ a divisor of $p-1$.  Then hidden conjugates
in $\Z_q \ltimes \Z_p$ are measurement reconstructible.
\end{theorem}
We then reduce the general problem of hidden subgroup reconstruction
in $\Z_q \ltimes \Z_p$ (and $A_p$) to
Theorem~\ref{thm:qhedralhcp}:
\begin{theorem}
  \label{thm:qhedralhsp} Let $p$ be prime and $q$ a divisor of $p-1$.
  The subgroups of the $q$-hedral groups $\Z_q \ltimes \Z_p$ are
  measurement reconstructible. In particular, the subgroups of the
  affine groups $A_p = \Z_{p-1}^* \ltimes \Z_p$ are measurement
  reconstructible.
\end{theorem}
In Theorems~\ref{thm:qhedralhcp} and \ref{thm:qhedralhsp} we give an
explicit set of efficiently computable measurements from which the
subgroup can be reconstructed, with a (possibly exponential) amount of
classical computation.

Finally, we show that the set of groups for which the HSP can be
solved in polynomial time has the following closure property:
\begin{theorem}
  \label{thm:semik}
  Let $H$ be a group for which hidden subgroups are fully
  reconstructible, and $K$ a group of polynomial size in $\log |H|$.
  Then hidden subgroups in any extension of $K$ by $H$, i.e.\ any
  group $G$ with $K \lhd G$ and $G/K \cong H$, are fully reconstructible.
\end{theorem}
This subsumes the results of~\cite{HallgrenRT00} on Hamiltonian
groups, and also those of~\cite{IvanyosMS01} on groups with commutator
subgroups of polynomial size.

\paragraph{The Non-Abelian Fourier Transform.} To solve the HSP for
the non-Abelian groups discussed above, we shall consider the more
general setting of non-Abelian Fourier analysis. Briefly, we treat a
representation as a homomorphism $\rho: G \to \U(d)$, where
$\U(d)$ denotes the group of unitary operators on $\C^d$. We call $d_\rho
= d$ the {\em dimension} of $\rho$. For a function $f: G \to \C$ and an
irreducible representation $\rho$, we let $\hat{f}(\rho)$ denote the Fourier
transform of $f$ at $\rho$, given by
\[
\hat{f}(\rho) = \sqrt{\frac{d_\rho}{|G|}} \,\sum_g f(g)\rho(g).
\]
A more complete description of the representations of a group $G$ and
the associated transform appear in Appendix~\ref{appendix:Fourier}.
The Fourier transform of a function of the form~\eqref{eq:super} is
then
\[
\hat{f}(\rho) = \sqrt{\frac{d_\rho}{|G||H|}} 
                \,\rho(c) \cdot \sum_{h \in H} \rho(h).
\]
As $H$ is a subgroup, $\sum_h \rho(h)$ is $|H|$ times a projection operator
(see, e.g., \cite{HallgrenRT00}); we write $\sum_h \rho(h) = |H| \,\pi_H$.
(Its rank is determined by the number of copies of the trivial
representation in the representation $\Ind_H^G {\mathbf 1}$.)  With
this notation, we write $\hat{f}(\rho) = \sqrt{n_\rho} \,\rho(c) \cdot \pi_H$ where
$n_\rho = d_\rho |H|/|G|$. For a $d \times d$ matrix $M$, we let $\norm{M}$
denote the matrix norm given by $\norm{M}^2 = \sum_{ij} \abs{M_{ij}}^2$.
Then the probability that we observe the representation $\rho$ is
$$
\norm{\hat{f}(\rho)}^2 = \norm{\sqrt{n_\rho} \,\rho(c) \pi_H}^2 = n_\rho \norm{\rho(c)}^2
\norm{\pi_H}^2 = n_\rho \,\rank \pi_H,
$$
where $\rank \pi_H$ is the rank of the projection operator $\pi_H$.
See~\cite{HallgrenRT00} for discussion.

\section{The Affine Group $A_p$}
Let $A_p$ be the affine group of size $p(p-1)$ for $p$ prime,
consisting of functions $(a,b): x \mapsto ax + b$ on $\Z_p$ acting by
composition, where $a \in \Z_p^*$ and $b \in \Z_p$.  Thus $A_p$ is a
semidirect product $\Z_p^* \ltimes \Z_p$ where $(a_1,b_1) \cdot
(a_2,b_2) = (a_1 a_2,b_1 + a_1 b_2)$ (we adopt the convention that
functions compose on the right). We enumerate the subgroups below:
\begin{itemize}
\item Let $N \cong \Z_p$ be the normal subgroup of size $p$ consisting
of elements of the form $(1,b)$.
\item Let $H$ be the non-normal subgroup of size $p-1$ consisting of
the elements of the form $(a,0)$.  Its conjugates $H^b = (1,b) \cdot H
\cdot (1,-b)$ consist of elements of the form $(a,(1-a)b)$. (In the
action on $\Z_p$, $H^b$ is the stabilizer of $b$).
\item More generally, if $a \in \Z_p^*$ has order $q$, let $N_a \cong
\Z_q \ltimes \Z_p$ be the normal subgroup consisting of all elements
of the form $(a^t,b)$, and let $H_a$ be the non-normal subgroup $H_a =
\langle (a,0) \rangle$ of size $q$.  Then $H_a$ consists of the
elements of the form $(a^t,0)$ and its conjugates $H_a^b=(1,b) \cdot
H_a \cdot (1,-b)$ consist of the elements of the form
$(a^t,(1-a^t)b)$.
\end{itemize}

To discuss $A_p$'s representations, fix a generator $\gamma$ of $\Z_p^*$
and let $\phi: \Z_p^* \to \Z_{p-1}$ be the isomorphism $\phi(\gamma^t) = t$. Let
$\omega_p$ denote the $p$'th root of unity $\e^{2 \pi i/p}$.  Then $G$ has
$p-1$ one-dimensional representations $\sigma_s$ which are simply the
representations of $\Z_p^* \cong \Z_{p-1}$ given by $\sigma_t((a,b)) =
\omega_{p-1}^{t \phi(a)}$ and one $(p-1)$-dimensional representation $\rho$.  In
the \emph{multiplicative} basis whose indices $j,k$ are elements of
$\Z_p^*$, we have:
\[ \rho((a,b))_{j,k} = \left\{ \begin{array}{ll}
   \omega_p^{bj}  & k = aj \bmod p \\
   0              & \mbox{otherwise}
   \end{array} \right. ,
\; 1 \leq j,k < p
\enspace .
\]
\remove{Alternately, in the {\em additive} basis whose indices $s,t$ are
elements of $\Z_{p-1}$ (i.e.\ exponents of the generator $\gamma$), we
have:
\[ \rho( (a,b) )_{s,t} = \left\{ \begin{array}{ll}
   \omega_p^{b \phi^{-1}(s)}  &  t = s + \phi(a) \bmod (p-1) \\
   0                          &  \mbox{otherwise}
   \end{array} \right. ,
\; 0 \leq s,t < p-1
\]}
We review the construction of these representations in
Appendix~\ref{appendix:Affine}.

The affine group --- and more generally, the $q$-hedral groups we
discuss below --- are {\em metacyclic} groups, i.e.\ extensions of a
cyclic group $\Z_p$ by a cyclic group $\Z_q$.  In~\cite{Hoyer97},
H{\o}yer showed how to perform the non-Abelian Fourier transform over
such groups in a polynomial (i.e.\ $\polylog(p)$) number of elementary
quantum operations.  (In fact, he does this only up to an overall
phase factor, but this is sufficient for our purposes.)

\paragraph{Conjugates of the Largest Non-Normal Subgroup.} In this
section we solve the Hidden Conjugate Problem, in which we are
promised that $f$ is a superposition over some coset of one of the
conjugates $H^b$ of the largest non-normal subgroup $H$, and our job
is to identify which conjugate, i.e.\ to identify $b$.  First note
that $n_\rho = d_\rho |H|/|G| = (p-1)/p = 1-1/p$.  Then a little calculation
shows that, in the multiplicative basis, $\pi(H^b)_{j,k} = (1/p-1)
\;\omega_p^{b(j-k)}$, $1 \leq j,k < p$. This is a circulant matrix of rank 1.
More specifically, every column is some root of unity times the vector
$(u_b)_j = (1/p-1) \;\omega_p^{bj}$, $1 \leq j < p$. This is also true of
$\rho(c) \cdot \pi(H^b)$; since $\rho(c)$ has one nonzero entry per column, left
multiplying by $\rho(c)$ simply multiplies each column of $\pi(H^b)$ by a
phase.  Therefore, we can first carry out a partial measurement on the
columns, and then transform the rows by left-multiplying $\rho(cH)$ by
the quantum Fourier transform over $\Z_{p-1}$, $Q_{\ell,j} =
(1/p-1)\;\omega_{p-1}^{-\ell j}$.  We can now infer $b$ by measuring the
frequency $\ell$.  We observe a given value of $\ell$ with probability
\[
P(\ell)
= \left| \frac{1}{p-1} \sum_{j=1}^{p-1} \omega_p^{bj} \omega_{p-1}^{-\ell j} 
  \right|^2
= \frac{1}{(p-1)^2} \left| \sum_{j=1}^{p-1} \e^{2 i \theta j} \right|^2
= \frac{1}{(p-1)^2} \frac{\sin^2 (p-1) \theta}{\sin^2 \theta}
\]
where $\theta = \left( \frac{b}{p} - \frac{\ell}{p-1} \right) \pi$.
Now note that for any $b$ there is an $\ell$ such that $|\theta| \leq
\pi/(2(p-1))$.  Since $(2x/\pi)^2 \leq \sin^2 x \leq x^2$ for $|x| \leq \pi/2$, this
gives $P(\ell) \geq (2/\pi)^2$.

Finally, the probability that we observed the $(p-1)$-dimensional
representation $\rho$ in the first place is $n_\rho = 1-1/p$. Thus if we
measure $\rho$, the column, and then $\ell$ and then guess that $b$
minimizes $|\theta|$, we will be right $\Omega(1)$ of the time.  We boost this
to high probability by repeating a polynomial number of times.

\paragraph{Subgroups with Large Index.} We focus next on the Hidden
Conjugate Problem for the subgroups $H_a$ where $a$'s order $q$ is a
proper divisor of $p-1$.  Recall that a given conjugate of $H_a$
consists of the elements of the form $(a^t,(1-a^t)b)$.  Then in the
multiplicative basis we have
\[ 
\pi(H_a^b)_{j,k} = \frac{1}{q} \left\{ \begin{array}{ll}
   \omega_p^{b(j-k)}  & k = a^t j \bmod p \mbox{ for some } t \\
   0              & \mbox{otherwise}
   \end{array} \right. ,
\; 1 \leq j,k < p
\enspace .
\]
In other words, the nonzero entries are those for which $j$ and $k$
are in the same coset of $\langle a \rangle \subset \Z_p^*$.  The rank of this
projection operator is thus the number of cosets, which is the index
$(p-1)/q$ of $\langle a \rangle$ in $\Z_p^*$.  Since $n_\rho$ is now $q/p$,
we again observe $\rho$ with probability $n_\rho \,\rank \pi(H) = (p-1)/p = 1-1/p$.

We will show that we can reconstruct the conjugates of $H_a$ in
polynomial time if $a$ has large order, in particular when the index
of $\langle a \rangle$ is $\polylog(p)$.  If $q$ is prime then $H_a$ is the only
non-normal subgroup of $\Z_q \ltimes \Z_p$, so we can completely solve
the Hidden Subgroup Problem for these groups.  For instance, if $q$ is
a {\em Sophie Germain} prime, i.e.\ one for which $2q+1$ is also a
prime, we can solve the HSP for $\Z_q \ltimes \Z_{2q+1}$. \textsl{This
  establishes Theorem~\ref{thm:hsp}}.

Following the same procedure as before, we do a partial measurement on
the columns of $\rho$, and then Fourier transform the rows.  After
changing the variable of summation from $t$ to $-t$ and adding a phase
shift of $\e^{-i \theta (p-1)}$ inside the $|\cdot|^2$, the probability we
observe a frequency $\ell$, assuming we find ourselves in the $k$'th
column, is
\begin{equation}
\label{eq:other}
P(\ell) = 
  \left| \frac{1}{\sqrt{q(p-1)}} 
  \,\sum_{t=0}^{q-1} \omega_p^{bka^t} \omega_{p-1}^{-\ell a^t k}
  \right|^2
= \frac{1}{q(p-1)} \left| 
  \sum_{t=0}^{q-1} e^{i \theta (2 a^t k-(p-1))} 
  \right|^2
\enspace .
\end{equation}
Now note that the terms in the sum are of the form $e^{i \phi}$ where
(assuming w.l.o.g.\ that $\theta$ is positive) $\phi \in [-\theta (p-1),\theta (p-1)]$.
If we again take $\ell$ so that $|\theta| \leq \pi/(2(p-1))$, then $\phi \in [-\pi/2,\pi/2]$
and all the terms in the sum have nonnegative real parts.  We will
lower bound the real part of the sum by showing that a constant
fraction of the terms have $\phi \in (-\pi/3,\pi/3)$, and thus have real part
more than $1/2$.  This is the case whenever $a^t k \in (p/6,5p/6)$, so
it is sufficient to prove the following lemma:

\begin{lemma}
  Let $a$ have order $q = p/\polylog(p)$.  Then for any $\eps > 0$ at
  least $(1/3-\eps) q$ of the elements in the coset $\langle a \rangle k$ are in
  the interval $(p/6,5p/6)$.
\end{lemma}

\noindent
{\bf Proof.}  We will prove this using {\em Gauss sums}, which
quantify the interplay between the additive and multiplicative
behavior of $\Z_p$ and thus establish bounds on the distribution of
powers of $a$.  Specifically, if $a$ has order $q$ in $\Z_p^*$ then
for any integer $k \not\equiv 0\; (\bmod p)$ we have $\sum_{t = 0}^{q - 1}
\omega_p^{a^t k} = \ord(p^{1/2}) = o(p)$. (See
Appendix~\ref{appendix:gauss-sums}.)

Now suppose $s$ of the elements $x$ in $\langle a \rangle k$ are in the set
$(p/6,5p/6)$, for which $\re \omega_p^x \geq -1$, and the other $q-s$ elements
are in $[0,p/6] \cup [5p/6,p)$, for which $\re \omega_p^x \geq 1/2$.  Thus we
have $\re \sum_{t = 0}^{q-1} \omega_p^{a^t k} \geq \,(q/2) - \,(3s/2)$. If $s \leq
(1/3-\eps) q$ for any $\eps > 0$ this is $\Theta(q)$, a
contradiction.  \hfill $\Box$

Now that we know that a fraction $1/3-\eps$ of the terms
in~\eqref{eq:other} have real part at least $1/2$ and the others have
real part at least $0$, we can take $\eps = 1/12$ (say) and write
\[
P(\ell) \geq \frac{1}{q(p-1)} \left( \frac{q}{8} \right)^2 
 = \frac{1}{8} \frac{q}{p-1} = \frac{1}{\polylog(p)}
\enspace .
\]
Thus we observe the correct frequency with polynomially small
probability, and we again boost this to high probability by repeating
a polynomial number of times. \textsl{This establishes Theorem~\ref{thm:hcp}.}

\section{The $q$-hedral Groups}

In general, if $a$ has multiplicative order $q$, then we are in the
subgroup $\Z_q \ltimes \Z_p \subset A_p$, the $q$-hedral group.  In this
section we show that the conjugates of $H_a$ are then measurement
reconstructible --- i.e.\ are information-theoretically
reconstructible from a polynomial number of quantum queries given by a
polynomial size quantum circuit, followed by a possibly exponential
amount of classical computation.  It follows that subgroups of the
$q$-hedral groups are measurement reconstructible
whenever $q$ has $\polylog(p)$ divisors --- for instance, $A_p$ (where $q=p-1$)
if $p$ is a Fermat prime $2^k + 1$. (Note also that for a
prime selected at random in $\{1, \ldots, n\}$ for large $n$, $p-1$ has no
more than $\polylog(p)$ divisors with high probability.) This
generalizes the results of Ettinger and H{\o}yer~\cite{EttingerH98} who
showed this for the case $q=2$, i.e.\ the dihedral groups.

The representations of $\Z_q \ltimes \Z_p$ include the $q$
one-dimensional representations of $\Z_q$ given by $\sigma_\ell((a^t,b)) =
\omega_q^{\ell t}$, $\ell \in \Z_q$ and $(p-1)/q$ $q$-dimensional representations
$\rho_k$,
\[ \rho_k(a^u,b) )_{s,t} = \left\{ \begin{array}{ll}
   \omega_p^{k a^s b}  &  t = s + u \bmod q \\
   0                     &  \mbox{otherwise}
   \end{array} \right. ,
\; 0 \leq s,t < q
\enspace .
\]
Here $k$ ranges over the elements of $\Z_p^* / \Z_q$, or, to put it
differently, $k$ takes values in $\Z_p^*$ but $\rho_k$ and
$\rho_{k'}$ are isomorphic if $k$ and $k'$ are in the same coset of
$\langle a \rangle$.  These $\rho_k$ are simply the $(p-1)/q$ diagonal
blocks of the $(p-1)$-dimensional representation $\rho$ of $A_p$ (this
is perhaps a little easier to see in the additive basis).

\begin{sloppypar}
Then summing $\rho_k$ over the elements $(a^t,(1-a^t)b)$ gives
$\pi_k(H_a^b)_{s,t} = (1/q) \; \omega_p^{k(a^s-a^t)b}$, $0 \leq s,t < q$.
This is again a matrix of rank 1, where each column (even after left
multiplication by $\rho_k(c)$) is some root of unity times the vector $(u_k)_s =
(1/q) \;\omega_p^{k a^s b}$. Note that $n_\rho = q/p$.
\end{sloppypar}

We now wish to show that there is a measurement whose outcomes given
two distinct values of $b$ have polynomial total variation distance.
First, we perform a series of partial measurements as follows:
\textsl{(i.)} measure the name of the representation; \textsl{(ii.)}
measure the column of the representation; \textsl{(iii.)} perform a
POVM with $q$ outcomes, in each of which $s$ is $u$ or $u+1 \bmod q$
for some $u \in \Z_q$.
The total probability we observe one of the $q$-dimensional representations,
since there are $(p-1)/q$ of them, is $n_\rho (p-1)/q = 1-1/p$.  
Then these three partial measurements determine $k$, remove the
effect of the coset, and determine that $s$ has one of two values, $u$
or $u+1$.  Up to an overall phase we can write this as a
two-dimensional vector
\[ \frac{1}{\sqrt{2}}
  \ve
  \omega_p^{k a^u b} \\
  \omega_p^{k a^{u+1} b}
  \ctor
\]
We now apply the Hadamard transform $(1/\sqrt{2}) {\scriptsize \mat 1
  & \!\!\!1 \\ 1 & \!\!\!\!-1 \rix}$ and measure $s$.  The probability
we observe $u$ and $u+1$ is then $\cos^2 \theta$ and $\sin^2 \theta$
respectively, where $\theta = (\pi k a^u (a-1) b)/p$.  Now when we observe a
$q$-dimensional representation, the $k$ we observe is uniformly
distributed over $\Z_p^* / \Z_q$, and when we perform the POVM, the
$u$ we observe is uniformly distributed over $\Z_q$.  It follows that
the coefficient $m = k a^u (u-1)$ is uniformly distributed over
$\Z_p^*$.  For any two distinct $b$, $b'$, the total variation
distance is then
\begin{eqnarray*}
& & \frac{1}{2(p-1)} \sum_{m \in \Z_p^*} 
  \left( \left| \cos^2 \frac{\pi m b}{p} - \cos^2 \frac{\pi m b'}{p} \right|
  + \left| \sin^2 \frac{\pi m b}{p} - \sin^2 \frac{\pi m b'}{p} \right|
  \right) \\
& = & \frac{1}{p-1} \sum_{m \in \Z_p} 
  \left| \cos^2 \frac{\pi m b}{p} - \cos^2 \frac{\pi m b'}{p} \right| 
\;\;=\;\; \frac{1}{2(p-1)} \sum_{m \in \Z_p}
  \left| \cos \frac{2 \pi m b}{p} - \cos \frac{2 \pi m b'}{p} \right| \\
& \geq & \frac{1}{4(p-1)} \sum_{m \in \Z_p}
  \left( \cos \frac{2 \pi m b}{p} - \cos \frac{2 \pi m b'}{p} \right)^2   
\;\;=\;\; \frac{p}{4(p-1)} \;\;>\;\; \frac{1}{4} \enspace .
\end{eqnarray*}
(Adding the $m=0$ term contributes zero to the sum in the second line.
In the third line we use the facts that $|x| \leq x^2/2$ for all $|x|
\leq 2$, the average of $\cos^2$ is $1/2$, and the two cosines have
zero inner product.)

\begin{sloppypar}
Since the total variation distance between any two distinct conjugates
is bounded below by a constant, by standard results in probability
theory we can distinguish between the $p$ different conjugates with
only $\ord(\log p) = \poly(n)$ queries.  \textsl{Thus hidden conjugates in
$q$-hedral groups are measurement reconstructible, completing the
proof of Theorem~\ref{thm:qhedralhcp}.}
\end{sloppypar}

What remains to be seen is that in a group of form $\Z_q \ltimes
\Z_p$, where $q \mid p-1$, it is possible to determine the order of a
hidden subgroup. Were this possible, based on
Theorem~\ref{thm:qhedralhcp}, we could (measurement) reconstruct
arbitrary hidden subgroups of $\Z_q \ltimes \Z_p$. Let $H$ be a hidden
subgroup of $\Z_q \ltimes \Z_p$ given by the oracle $f: \Z_q \ltimes
\Z_p \to S$, and let $p_1^{\alpha_1}\ldots p_k^{\alpha_k}$ be the prime
factorization of $q$, in which case $k \leq \sum_i \alpha_i = \ord(\log q)$.
For each $i \in [k]$, we will determine if $p_i^{\alpha_i} \mid |H|$. This
suffices to determine $|H|$, at which point the subgroup $H$ can be
determined by Theorem~\ref{thm:qhedralhcp}.

By initially applying the techniques of \cite{HallgrenRT00} (the weak
standard method), we may (fully) reconstruct $H$ if $H$ is a
non-trivial normal subgroup. (This follows because these particular
semidirect product groups have the special property that if $A$ is a
non-trivial normal subgroup and $A \subset B$, then $B$ is normal; in
particular, the normal core
$$
\bigcap_{\gamma \in  G} \gamma C \gamma^{-1}
$$
of any non-normal subgroup $C$ is the identity group.) It remains
to consider non-normal subgroups $H$. Recall that in this case, $H$ is
cyclic and $|H|$ is equal to the order of $a$, where $H = \langle(a,b)\rangle$.
Now, for each $i \in [k]$ and $1 \leq \alpha \leq \alpha_i$, let $\Upsilon_i^\alpha: \Z_{q}
\ltimes \Z_p \to \Z_{q/p_i^{\alpha}}$ be the homomorphism given by
$$
\Upsilon_i^\alpha: (a,b) \mapsto a^{p_i^{\alpha}}.
$$
Then let $A_i^{\alpha_i} = \ker \Upsilon_i^\alpha = \{ \gamma \in \Z_q \ltimes \Z_p \mid
\gamma^{p_i^{\alpha_i}} = \id \}$, where $\id$ denotes the identity element of
$\Z_q \ltimes \Z_p$. $A_i^{\alpha_i}$ is the subgroup of $\Z_q \ltimes
\Z_p$ consisting of all elements whose orders are a multiple of
$p_i^{\alpha}$.  Consider now the function
$$
(f,\Upsilon_i^\alpha) : \Z_q \ltimes \Z_p \to S \times \Z_{q/p_i^\alpha}
$$
given by $(f,\Upsilon_i^\alpha)(\gamma) = (f(\gamma),\Upsilon_i^\alpha(\gamma))$. Observe that
$(f,\Upsilon_i^\alpha)$ is constant (and distinct) on the left cosets of $H \cap
A_i^{\alpha}$ and, furthermore, the subgroup $H \cap A_i^\alpha$ has order
$p^\alpha$ if and only if $p^\alpha$ divides the order of $a$. We may then
determine if $H \cap A_i^\alpha$ has order $p^\alpha$ by assuming that it does,
applying the result of Theorem~\ref{thm:qhedralhcp}, and checking the
result against the original oracle $f$. This allows us to determine
the prime factorization of $|H|$, as desired. \textsl{Therefore, all
  subgroups of the $q$-hedral groups $\Z_q \ltimes \Z_p$ are
  measurement reconstructible, completing the proof of
  Theorem~\ref{thm:qhedralhsp}.}

However, as in the dihedral case~\cite{EttingerH98}, we know of no
polynomial-time algorithm which can reconstruct the most likely $b$
from these queries.

\section{Failure of the Abelian Fourier Transform}

Suppose we try to reconstruct subgroups of $A_p$ using the Abelian
Fourier transform over the direct product $\Z_p^* \times \Z_p$ instead of
using $A_p$'s non-Abelian structure as a semidirect product.  We first
consider trying to solve the hidden conjugate problem for $H_a$ where
$a$ has order $p-1$.

If $a$ is a generator, the characters of $\Z_p^* \times \Z_p$ are simply
$\rho_{k,\ell}(a^t,b) = \omega_{p-1}^{kt} \omega_p^{\ell b}$.  Summing these over $H_a =
\{ (a^t, (1-a^t)b \}$ shows that we observe the character $(k,\ell)$ with
probability
\[ P(k,\ell)
  = \frac{1}{p \,(p-1)^2} \left| 
  \sum_{t \in Z/(p-1)} \omega_{p-1}^{kt} \omega_p^{\ell (1-a^t) b} \right|^2
  = \frac{1}{p \,(p-1)^2} \left| 
     \sum_{x \in \Z_p^*} \omega_{p-1}^{k \log_a x} \omega_p^{-\ell x b} 
    \right|^2
\enspace . 
\]
This is the inner product of a multiplicative character with an
additive one, which is another Gauss sum.  In particular, assuming $b
\neq 0$, we have $P(0,0) = 1/p$, $P(0, \ell \neq 0) = 1/ (p\,(p-1)^2)$, $P(k \neq
0, 0) = 0$, and $P(k \neq 0, \ell \neq 0) = 1/(p-1)^2$. (See
Appendix~\ref{appendix:gauss-sums}.) Since these probabilities don't
depend on $b$, the different conjugates $H_a^b$ with $b \neq 0$ are
indistinguishable from each other.  Thus it appears essential that we
use the use non-Abelian Fourier transform and the high-dimensional
representations of $A_p$.

(For the $q$-hedral groups, when $q$ is small enough it is
information-theoretically possible to reconstruct the subgroup from
the Abelian Fourier transform.  In fact, Ettinger and
H{\o}yer~\cite{EttingerH98} use the Abelian Fourier transform over $\Z_2
\times \Z_p$ in their reconstruction algorithm for the dihedral groups.)

\section{Closure Under Extending Small Groups}

In this section we prove Theorem~\ref{thm:semik}, that for any
polynomial-size group $K$ and any $H$ for which we can solve the HSP,
we can also solve the HSP for any extension of $K$ by $H$, i.e.\ any
group $G$ with $K \lhd G$ and $G/K \cong H$.  (Note that this is more
general than split extensions, i.e.\ semidirect products $H \ltimes
K$.)  This includes the case discussed in~\cite{HallgrenRT00} of
Hamiltonian groups, since all such groups are direct products (and
hence extensions) by Abelian groups of the quaternion group
$Q_8$~\cite{Rotman94}.  It also includes the case discussed
in~\cite{FriedlIMSS02} of groups with commutator subgroups of
polynomial size, such as extra-special $p$-groups, since in that case
$K=G'$ and $H \cong G/G'$ is Abelian.  Indeed, our proof is an easy
generalization of that in~\cite{FriedlIMSS02}.

We assume that $G$ and $K$ are encoded in such a way that
multiplication can be carried out in classical polynomial time.  We
fix some transversal $t(h)$ of the left cosets of $K$. First, note
that any subgroup $L \subseteq G$ can be described in terms of i) its
intersection $L \cap K$, ii) its projection $L_H = L/(L \cap K) \subseteq H$, and
iii) a representative $\eta(h) \in L \cap (t(h) \cdot K)$ for each $h \in L_H$.
Then each element of $L_H$ is associated with some left coset of $L \cap
K$, i.e.\ $ L = \bigcup_{h \in L_H} \eta(h) \cdot (L \cap K)$.  Moreover, if $S$ is a
set of generators for $L \cap K$ and $T$ is a set of generators for
$L_H$, then $S \cup \eta(T)$ is a set of generators for $L$.

We can reconstruct $S$ in classical polynomial time simply by querying
$F$ on all of $K$.  Then $L \cap K$ is the set of all $k$ such that $F(k)
= F(1)$, and we construct $S$ by adding elements of $L \cap K$ to it one
at a time until they generate all of $L \cap K$. 

To identify $L_H$, as in~\cite{FriedlIMSS02} we define a new function
$F'$ on $H$ consisting of the unordered collection of the values of
$F$ on the corresponding left coset of $K$: $F'(h) = \{ F(g) \,|\, g \in
t(h) \cdot K \}$. Each query to $F'$ consists of $|K| = \poly(n)$ queries
to $K$.  The level sets of $F'$ are clearly the cosets of $L_H$, so we
reconstruct $L_H$ by solving the HSP on $H$.  This yields a set $T$ of
generators for $L_H$.

It remains to find a representative $\eta(h)$ in $L \cap (t(h) \cdot
K)$ for each $h \in T$.  We simply query $F(g)$ for all $g \in t(h)
\cdot K$, and set $\eta(h)$ to any $g$ such that $F(g) = F(1)$.  Since
$|T| = \ord(\log |H|) = \poly(n)$ this can be done in polynomial time,
and we are done.

Unfortunately, we cannot iterate this construction more than a
constant number of times, since doing so would require a
superpolynomial number of queries to $F$ for each query of $F'$.  If
$K$ has superpolynomial size it is not clear how to obtain $\eta(h)$,
even when $H$ has only two elements: this is precisely the difficulty
with the dihedral group. \textsl{This completes the proof of Theorem~\ref{thm:semik}.}

\bigskip {\bf Acknowledgements.}  We are grateful to Wim van Dam,
Frederic Magniez, Martin R\"{o}tteler, and Miklos Santha for helpful
conversations, and to Sally Milius and Tracy Conrad for their support.
Support for this work was provided by the California Institute of
Technology's Institute for Quantum Information (IQI), the Mathematical
Sciences Research Institute (MSRI), the Institute for Advanced Study
(IAS), NSF grants ITR-0220070 and QuBIC-0218563, the Charles Lee
Powell Foundation, and the Bell Fund.

\appendix

\section{The Non-Abelian Fourier Transform}
\label{appendix:Fourier}

To solve the HSP for the non-Abelian groups discussed above, we shall
have to consider the more general setting of non-Abelian Fourier
analysis.  Here, instead of the familiar basis functions $h_k(x) =
\omega_p^{kx}$, which are homomorphisms from $\Z_p$ into $\C$, we have {\em
  representations} $\rho$ which are homomorphisms from $G$ into $\U(d)$,
the group of unitary $d \times d$ matrices with entries in $\C$. We call
$d_\rho = d$ the {\em dimension} of $\rho$.

We say that two representations $\rho: G \to \U(d)$ and $\sigma: G \to \U(d)$ are
\emph{isomorphic} if there is a non-singular linear map $\iota: \C^d \to
\C^d$ for which $\rho(g) \circ \iota = \iota \circ \sigma(g)$ for every $g \in G$. Though there
are an infinite number of non-isomorphic representations of a given
group $G$, there is a natural notion of ``decomposition'' that applies
to such representations; with respect to this notion, a finite group
$G$ has a finite number of ``irreducible'' representations up to
isomorphism, and every other representation may be expressed in terms
of these basic building blocks. Specifically, we say that a
representation $\rho: G \to \U(d)$ is \emph{reducible} if there is a
nontrivial subspace $\{ 0 \} \subsetneq W \subsetneq \C^d$ with the
property that $\rho(g)(W) \subset W$ for all $g \in G$. A representation is
\emph{irreducible} if no such subspace exists.

For a given group $G$, there are only a finite number of irreducible
representations upto isomorphism; we let $\hat{G}$ denote a set of
irreducible representations of $G$ containing one from each
isomorphism class.

Let $f: G \to \C$ be a function and $\rho$ an irreducible representation of
$G$. Then the \emph{Fourier transform of $f$ at $\rho$}, written
$\hat{f}(\rho)$, is the operator
\[
\hat{f}(\rho) = \sqrt{\frac{d_\rho}{|G|}} \,\sum_g f(g)\rho(g).
\]
The functional notation $\hat{f}(\rho)$ is somewhat misleading, as
$\hat{f}(\rho)$ is a $d_\rho \times d_\rho$ matrix, the dimension $d_\rho$ being
determined by the representation $\rho$. By selecting an orthonormal
basis for $\C^{d_\rho}$ for each $\rho$, we may associate with $f$ the
family of complex numbers $\hat{f}(\rho)_{ij}$, where $1 \leq i,j \leq d_\rho$;
With the constants $\sqrt{{d_\rho}/|G|}$, the linear transformation 
$$
f \mapsto \langle \hat{f}(\rho)_{i,j}\rangle_{\rho \in \hat{G}, 1 \leq i,j \leq d_\rho}
$$
is in fact unitary.

The Fourier transform of a function of the form~\eqref{eq:super} is
then
\[
\hat{f}(\rho) = \sqrt{\frac{d_\rho}{|G||H|}} 
                \,\rho(c) \cdot \sum_{h \in H} \rho(h).
\]
As $H$ is a subgroup, $\sum_h \rho(h)$ is $|H|$ times a projection operator
(see, e.g., \cite{HallgrenRT00}); we write $\sum_h \rho(h) = |H| \,\pi_H$.
(Its rank is determined by the number of copies of the trivial
representation in the representation $\Ind_H^G {\mathbf 1}$.)  With
this notation, we write $\hat{f}(\rho) = \sqrt{n_\rho} \,\rho(c) \cdot \pi_H$ where
$n_\rho = d_\rho |H|/|G|$. For a $d \times d$ matrix $M$, we let $\norm{M}$
denote the matrix norm given by $\norm{M}^2 = \sum_{ij} \abs{M_{ij}}^2$.
Then the probability that we observe the representation $\rho$ is
$$
\norm{\hat{f}(\rho)}^2 = \norm{\sqrt{n_\rho} \,\rho(c) \pi_H}^2 = n_\rho \norm{\rho(c)}^2
\norm{\pi_H}^2 = n_\rho \,\rank \pi_H,
$$
where $\rank \pi_H$ is the rank of the projection operator $\pi_H$.
See~\cite{HallgrenRT00} for more discussion.

\section{Constructing $A_p$'s Representations; Induced Representations}
\label{appendix:Affine}

In this Appendix we construct the $(p-1)$-dimensional representation
of $A_p$ by inducing upward from a one-dimensional representation of
the normal subgroup $N \cong \Z_p$. We begin with a short discussion of
induced representations.

Let $G$ be a group, $H$ a subgroup of $G$, and $\sigma: H \to
\U(d)$ a representation of $H$. We shall define a representation
$\Ind_H^G \sigma$ of $G$, the \emph{induced} representation.
Let $\Gamma = \{ \gamma_1, \ldots, \gamma_t\} \subset G$ be a left
transversal of $H$ in $G$, so
that $G = \cup_{\gamma \in \Gamma} \gamma H$, this union being disjoint.
The
representation $\Ind_H^G \sigma$ is defined on the vector space of
dimension $d |G| / |H|$ whose elements are formal sums $\sum_{\gamma \in
\Gamma} \gamma \cdot
v_\gamma$, where each $v_\gamma \in \C^d$. Addition and scalar
multiplication are
given by the rule $\sum \gamma \cdot u_\gamma + \sum \gamma \cdot
v_\gamma = \sum \gamma \cdot (u_\gamma + v_\gamma)$ and $c
\sum \gamma \cdot v_\gamma = \sum \gamma \cdot cv_\gamma$. Then
$\Ind_H^G \sigma$ is defined by linearly
extending the rule
$$
\left[\Ind_H^G \sigma(g)\right] \gamma \cdot v_\gamma \mapsto \gamma'
\cdot \sigma(h) v_\gamma
$$
where $(\gamma',h)$ is the unique pair in $\Gamma \times H$ so that $g
\gamma = \gamma' h$.

Returning now to the affine group, let $\tau_t(1,b) \mapsto
\omega_p^{tb}$ for $0 \leq
t < p$ be the $p$ distinct one-dimensional characters of the normal
subgroup $N = \Z_p$. Let $H = A_p / N \cong \Z_p^*$. Consider the
conjugation action of $H$ on these characters: that is, define $(a,0)
\odot \tau_t(1,b) = \tau_t [(a,0)(1,b)(a,0)^{-1}] = \tau_t(1, ab) =
\tau_{at}(1,b)$.
Note that this action has two orbits, one consisting of the trivial
character $\tau_0$ and the other consisting of all non-trivial
character.

Now, considering the first orbit, consisting of $\tau_0$ alone, we see
that the isotropy subgroup is all of $H$. Now, let $\rho_0$ be the
extension of $\sigma_0$ to all of $H$ (which makes sense, since it was
stable under the $H$-action). Then for each irreducible representation
$\check{\sigma}$ of $H$, we get an irreducible representation $\sigma =
\Ind_{HN}^{A_p} (\rho_0 \otimes \check{\sigma})$. (Note that this gives rise to the
representations $\sigma_s$ above.)

Focusing on the other orbit, for simplicity consider $\check{\sigma}_1$.
Since $H$ is cyclic, the isotropy subgroup of $\sigma_1$ is the identity
subgroup and this gives rise to the representation $\rho =
\Ind_{N}^{A_p}
\check{\sigma}_1$. Now $\Ind_N^{A_p}$ operates on the vector space $W =
(1,0)\C \oplus \ldots \oplus (p-1,0)\C$. The action is
\[
[\Ind_N^{A_p}(a,b)] \cdot (i,0) \mapsto \check{\sigma}_1((ai)^{-1}b)
(ai,0).
\]
so that
\[
[\Ind_N^{A_p}(a,b)]_{j,k} = \left\{ \begin{array}{ll}
   \omega_p^{bj}  & k = aj \bmod p \\
   0              & \mbox{otherwise}
   \end{array} \right. ,
\; 1 \leq j,k < p
\]
which is precisely the $(p-1)$-dimensional representation $\rho$ in the
multiplicative basis.  We can construct the $q$-dimensional
representations of the $q$-hedral groups in a similar way.

\section{Notes on Exponential Sums}
\label{appendix:gauss-sums}

The basic \emph{Gauss sum} bounds the inner products of additive and
multiplicative characters of $\F_p$, the finite field with $p$
elements. Definitive treatments appear in~\cite[{\S}5]{LidlN97}
and~\cite{KonyaginS99}. Considering $\F_p$ as an additive group with
$p$ elements, we have $p$ additive characters $\chi_s : \F_p \to \C$, for
$s \in \F_p$, given by
$$
\chi_s : z \mapsto \omega_p^{sz},
$$
where $\omega_p = e^{2 \pi i/p}$ is a primitive $p$th root of unity.
Likewise considering the elements of $\F_p^* = \F_p \setminus \{ 0\}$ as a
multiplicative group, we have $p-1$ characters $\psi_t : \F_p^* \to \C$,
for $t \in \F_p^*$, given by
$$
\psi_t : g^z \mapsto \omega_{p-1}^{tz},
$$
where $\omega_{p-1} = e^{2 \pi i/(p-1)}$ is a primitive $p-1$st root of
unity and $g$ is a multiplicative generator for the (cyclic) group
$\F_p^*$.

With this notation the basic Gauss sum is the following:
\begin{theorem}
  Let $\chi_s$ be a multiplicative character and $\psi_t$ an additive
  character of $\F_p$.  If $s \neq 0$ and $t \neq 1$ then 
  $$
  \Bigl | \sum_{z \in \F_p^*} \chi_s(z)\psi_t(z) \Bigr | = \sqrt{p}.
  $$
  Otherwise
  $$
  \sum_{z \in \F_p^*} \chi_s(z) \psi_t(z) = \begin{cases}
    p-1 & \text{if}\; s = 0, t = 1,\\
    -1  & \text{if}\; s = 0, t \neq 1,\\
    0   & \text{if}\; s \neq 0, t = 1.\\
  \end{cases}
  $$
\end{theorem}
See~\cite[{\S}5.11]{LidlN97} for a proof.

This basic result has been spectacularly generalized. In the body of
the paper we require bounds on additive characters taken over
multiplicative subgroups of $\F_p^*$. Such sums are discussed in detail
in \cite{KonyaginS99}. The specific bound we require is the following.

\begin{theorem}
  Let $\chi_t$ be a nontrivial additive character of $\F_p$ and $a \in
  \F_p^*$ an element of multiplicative order $q$. Then 
  $$
  \sum_{z = 0}^{q - 1} \chi_t(a^z) = \begin{cases}
    \ord(p^{1/2}), &\text{if}\;q \geq p^{2/3},\\
    \ord(p^{1/4} q^{3/8}), &\text{if}\;p^{1/2} \leq q \leq p^{2/3},\\
    \ord(p^{1/8} q^{5/8}), &\text{if}\;p^{1/3} \leq q \leq p^{1/2}.
  \end{cases}
  $$  
\end{theorem}
See \cite[{\S}2]{KonyaginS99} for a proof. 

Note that in the body of the paper, we use $\Z_p$ to denote the
additive group of integers modulo $p$ and $\Z_p^*$ to denote the
multiplicative group of integers modulo $p$.
\end{document}